\documentclass[12pt]{article}
\usepackage{graphicx}
\def \b{{\cal B}} 
\def \bea{\begin{eqnarray}}
\def \beq{\begin{equation}}
\def \eea{\end{eqnarray}}
\def \eeq{\end{equation}}

\def \st{\sqrt{3}} 
\def \sx{\sqrt{6}}

\begin{document} 
\rightline{EFI 09-06} 
\rightline{arXiv:0903.1796} 
\rightline{March 2009} 
\bigskip
\centerline{\bf MESON-PHOTON TRANSITION FORM FACTORS}
\centerline{\bf IN THE CHARMONIUM ENERGY RANGE} 
\bigskip 
 
\centerline{Jonathan L. Rosner\footnote{rosner@hep.uchicago.edu}} 
\centerline{\it Enrico Fermi Institute and Department of Physics} 
\centerline{\it University of Chicago, 5640 S. Ellis Avenue, Chicago, IL 60637} 
\begin{quote}
The study of electron-positron collisions at center-of-mass energies
corresponding to charmonium production has reached new levels of sensitivity
thanks to experiments by the BES and CLEO Collaborations.  Final states
$\gamma P$, where $P$ is a pseudoscalar meson such as $\pi^0$, $\eta$, and
$\eta'$ can arise either from charmonium decays or in the continuum through a
virtual photon: $e^+ e^- \to \gamma^* \to \gamma P$.  Estimates of this
latter process are given at center-of-mass (c.m.) energies corresponding to
the $J/\psi(1S)$, $\psi(2S)$, and $\psi(3770)$ resonances.
\end{quote}
 
\leftline{PACS numbers: 13.40.Gp, 13.66.Bc, 14.40.Aq, 14.40.Cs} 
\bigskip 
 
\centerline{\bf I.  INTRODUCTION} 
\bigskip 

The decays of charmonium states to a photon and a neutral pseudoscalar meson
$P=(\pi^0,\eta,\eta')$ can shed light on a number of different mechanisms,
including two-gluon couplings to $q \bar q$ states, vector meson dominance,
mixing of heavy quarkonium with bound states of light quarks and antiquarks,
and final-state radiation by light quarks \cite{CLEOGP}.  In these final states
there can be direct contributions from the continuum process
\beq \label{eqn:cont}
e^+ e^- \to \gamma^* \to \gamma P~,
\eeq
where $P$ is a neutral pseudoscalar meson.  The $\gamma^*$--$\gamma$--$P$
vertex was shown \cite{Brodsky:1981rp} to be characterized by
a form factor $F(Q^2)$, where $Q^2 \equiv - q^2$ and $q$ is the four-momentum
of the virtual photon $\gamma^*$, behaving as $F(Q^2) \to 2 f_\pi / Q^2$.
Here $f_\pi = 93$ MeV is the neutral-pion decay constant.  In the present
article we evaluate the cross sections for $P=\pi^0$, $\eta$, and $\eta'$ at
$\sqrt{s} = 3097,~3696$, and 3773 MeV.  These results are relevant to
possible continuum backgrounds in searches for decays of $J/\psi,~\psi(3686)$,
and $\psi(3770)$ to $\gamma P$, currently undertaken by the CLEO Collaboration
\cite{CLEOGP}.

In Section II we calculate the cross section for the process (\ref{eqn:cont}),
given the form factor estimate in Ref.\ \cite{Brodsky:1981rp}.  We then
estimate bounds on the actual cross section in Section III by comparing the
expected result for $e^+ e^- \to J/\psi \to \gamma \pi^0$ using vector
dominance with the observed (slightly larger) cross section.  We summarize in
Section IV.
\bigskip

\centerline{\bf II.  CONTINUUM $\gamma P$ PRODUCTION}
\bigskip

The $\gamma^*$--$\gamma$--$P$ vertex was discussed in Ref.\
\cite{Brodsky:1981rp}, where a proposal was made to test it in photon-photon
collisions with one photon highly off-shell.  In that case, the off-shell
photon will have a spacelike $Q^2 >0$.  Experimental tests in this regime were
indeed performed in Ref.\ \cite{Gronberg:1997fj}.  However, the reaction
(\ref{eqn:cont}) is equally useful in testing that vertex for timelike $Q^2 =
- q^2 < 0$.  The BaBar Collaboration \cite{Aubert:2006cy} has measured the
cross sections for $e^+ e^- \to \gamma \eta^{(\prime)}$ at $q^2 = 112$,
GeV$^2$ providing valuable information on the $\gamma^*$--$\gamma$--$\eta^
{(\prime)}$ vertex in the asymptotic regime.  We shall compare our predictions
with their results presently.

The vertex of interest has the form
\beq
\Gamma_\mu = - i e^2 F_{\pi^0}(Q^2) \epsilon_{\mu \nu \rho \sigma}
p^\nu \epsilon^\rho q^\sigma~.
\eeq
Here the form factor $F_{\pi^0}(Q^2)$ is expected to behave for large
$Q^2$ as \cite{Brodsky:1981rp}
\beq
F_{\pi^0}(Q^2) \to 2 f_\pi/Q^2~.
\eeq
The four-momentum of the pseudoscalar meson (here, $\pi^0$) is denoted as $p$,
while $\epsilon^\rho$ is the polarization vector of the outgoing on-shell
photon.

The differential cross section with respect to $\cos \theta$, where $\theta$ is
the angle the outgoing photon makes with the beam axis in the $e^+ e^-$
center-of-mass system, is
\beq \label{eqn:diff}
\frac{d\sigma(e^+ e^- \to \gamma^* \to \gamma \pi^0)}{d(\cos \theta)}
= \alpha^3 \left( \frac{\pi f_\pi}{s} \right)^2 K_{\pi^0}^3 (1+\cos^2 \theta)~,
~~K_P \equiv 1 - \frac{M_P^2}{s}~,
\eeq
where $s = q^2$ is the square of the center-of-mass energy and $M_P$ is the
mass of the pseudoscalar meson (here, $\pi^0$).  Integrating with respect to
$\cos \theta$, we find
\beq \label{eqn:tot}
\sigma(e^+ e^- \to \gamma^* \to \gamma \pi^0) = \frac{8 \alpha^3}{3} \left(
\frac{\pi f_\pi}{s} \right)^2 K_{\pi^0}^3~.
\eeq

This may be compared with the cross section for muon pair production
(neglecting $m_\mu$),
\beq
\sigma(e^+ e^- \to \gamma^* \to \mu^+ \mu^-) = \frac{4 \pi \alpha^2}{3 s}
= \frac{86.8~{\rm nb}}{s~({\rm GeV}^2)},
\eeq
which is 9.05 nb at $\sqrt{s} = 3.097$ GeV.  Thus
\beq \label{eqn:rpi1ss}
R^{\pi^0}_{\sigma}(s) \equiv\frac{\sigma(e^+ e^- \to\gamma^* \to \gamma \pi^0)}
     {\sigma(e^+ e^- \to \gamma^* \to \mu^+ \mu^-)}
 = \frac{2 \pi \alpha f_\pi^2}{s} K_{\pi^0}^3
\eeq
which is $4.11 \times 10^{-5}$ at $\sqrt{s} = 3.097$ GeV.  At this energy
we thus have
\beq \label{eqn:pi1s}
\sigma(e^+ e^- \to \gamma^* \to \gamma \pi^0) = 372~{\rm fb}~.
\eeq

The ratio (\ref{eqn:rpi1ss}) can also be used to predict the virtual-photon
contribution to the branching fraction $\b(J/\psi \to \gamma \pi^0)$ in
terms of $\b(J/\psi \to \mu^+ \mu^+)$:
\beq \label{eqn:rpi1sb}
R^{\pi^0}_{\b(J/\psi)} \equiv \frac{\b(J/\psi \to \gamma^* \to \gamma \pi^0)}
     {\b(J/\psi \to \gamma^* \to \mu^+ \mu^-)}
 = R^{\pi^0}_{\sigma}(M^2_{J/\psi})~.
\eeq
With $\b(J/\psi \to \mu^+\mu^-)$ assumed to be equal to $\b(J/\psi \to e^+e^-)
= 5.94\%$ \cite{PDG}, this implies
$\b(J/\psi \to \gamma^* \to \gamma \pi^0) = 2.44 \times 10^{-6}$, far below the
observed value \cite{PDG} of $(3.3^{+0.6}_{-0.4}) \times 10^{-5}$.  We shall
see in the next section that the vector-dominance process $J/\psi \to
\rho^{0*} \pi^0 \to \gamma \pi^0$ accounts for most if not all of the observed
branching fraction.

Whereas the asymptotic form factor $F_{\pi^0}(Q^2) = 2 f_\pi/Q^2$ of Ref.\
\cite{Brodsky:1981rp} implies $R^{\pi^0}_{\b(J/\psi)} = 4.11 \times 10^{-5}$,
a slightly larger value of $R^{\pi^0}_{\b(J/\psi)} = 10^{-4}$ is implied by
the calculation of Ref.\ \cite{Chernyak:1983ej}.  This would imply
\beq \label{eqn:pi1scz}
\sigma(e^+ e^- \to \gamma^* \to \gamma \pi^0) = 905~{\rm fb}~.
\eeq
for the continuum contribution at $\sqrt{s} = 3.097$ GeV.

We shall now scale the cross section estimates (\ref{eqn:pi1s}) and
(\ref{eqn:pi1scz}) to other energies and particle types.  Let the quark
model wave function of a neutral particle $P$ be characterized by the sum
of pairs $q_i \bar q_i$ with coefficients $c^P_i$:
\beq
|P \rangle = \sum_i c^P_i | q_i \bar q_i \rangle~.
\eeq
Then the ratio of the $\gamma^*$--$\gamma$--$P$ form factor to that
for $\pi^0$ is just
\beq
F_{P}(Q^2)/F_{\pi^0}(Q^2) = \sum_i c^P_i Q_i^2/ \sum_i c^{\pi^0}_i Q_i^2~.
\eeq
The $\eta$ and $\eta'$ may be represented as octet-singlet mixtures,
\beq
\eta = \cos \theta \eta_8 + \sin\theta \eta_1~,~~
\eta' = - \sin \theta \eta_8 + \cos \theta \eta_1~,
\eeq
\beq
\eta_8 \equiv (u \bar u + d \bar d - 2 s \bar s)/\sx~,~~
\eta_1 \equiv (u \bar u + d \bar d + s \bar s)/\st~.
\eeq
An approximate form which fits the data well \cite{etamix}, and which we shall
take in what follows, is
\beq \label{eqn:etamix}
\eta = (u \bar u + d \bar d - s \bar s)/\st~,~~
\eta' = (u \bar u + d \bar d + 2 s \bar s)/\sx~,
\eeq
corresponding to a mixing angle $\theta = -{\rm arcsin}(1/3)= -19.5^\circ$.
This implies \\ $|F_{\eta}(Q^2)/F_{\pi^0}(Q^2)|^2 = 32/27$, $|F_{\eta'}(Q^2)/
F_{\pi^0}(Q^2)|^2 = 49/32$.  (For comparison, the unmixed octet and singlet
states give $|F_{\eta_8}(Q^2)/F_{\pi^0}(Q^2)|^2 = 1/3$, $|F_{\eta_1}(Q^2)/
F_{\pi^0}(Q^2)|^2 = 8/3$.)

The above mixing {\it ansatz} implies
\bea
{\rm lim}_{q^2 \to \infty} |q^2 F_{\eta}(Q^2)| &=& 2 f_\pi\sqrt{\frac{32}{27}}
 = 202~{\rm MeV}~, \\
{\rm lim}_{q^2 \to \infty}|q^2 F_{\eta'}(Q^2)| &=& 2 f_\pi\sqrt{\frac{49}{27}}
 = 251~{\rm MeV}~,
\eea
in satisfactory agreement with the results from BaBar \cite{Aubert:2006cy} at
$q^2 = 112$ GeV$^2$:
\bea
|q^2 F_{\eta}(Q^2)| & = & 229 \pm 30 \pm 8~{\rm MeV}~,\\
|q^2 F_{\eta'}(Q^2)| & = & 251 \pm 19 \pm 8~{\rm MeV}~.
\eea
Other {\it ans\"atze} for $\eta$--$\eta'$ mixing have been explored, for
example, in Ref.\ \cite{Feldmann:1998vh}.  The two-gluon components of $\eta,
\eta'$ are not considered here, but are treated in Ref.\ \cite{Kroll:2002nt}.

The kinematic factors $K^3$ are summarized for $P = \pi^0,~\eta, \eta'$ and
c.m.\ energies corresponding to $J/\psi$, $\psi(2S)$, and $\psi(3770)$ in Table
\ref{tab:K}.  Also shown are the continuum cross sections at these c.m.\
energies.  At 3.77 GeV, the predicted continuum cross sections for $e^+ e^-
\to \gamma \eta^{(\prime)}$ are (0.19,0.25) pb, consistent with the values of
$(0.17^{+0.05}_{-0.04}\pm0.03,~0.21^{+0.07}_{-0.05}\pm0.03)$ pb observed
by CLEO \cite{CLEOGP}.  Thus, it is consistent to assume that the $\gamma \eta$
and $\gamma \eta'$ signals at 3.77 GeV come entirely from continuum.  Upper
bounds on $\b[\psi(3770) \to \gamma \eta^{(\prime)}]$ under various scenarios
of interference between direct decay and interference are quoted in Ref.\
\cite{CLEOGP}.

\begin{table}
\caption{Kinematic suppression factors $K^3 \equiv [1 - (M_P^2/s)]^3$ and
continuum cross sections $\sigma(e^+ e^- \to \gamma^* \to \gamma P)$ for
neutral pseudoscalar mesons $P$.
\label{tab:K}}
\begin{center}
\begin{tabular}{c c c c c c c} \hline \hline
 & \multicolumn{3}{c}{$K^3$} & \multicolumn{3}{c}{$\sigma$ (fb)} \\
    $P$   & $J/\psi$ & $\psi(2S)$ & $\psi(3770)$ &
 $J/\psi$ & $\psi(2S)$ & $\psi(3770)$ \\ \hline
 $\pi^0$ & 0.994 & 0.996 & 0.996 & 372 & 186 & 169 \\
 $\eta$  & 0.909 & 0.935 & 0.938 & 403 & 207 & 189 \\
 $\eta'$ & 0.740 & 0.811 & 0.819 & 502 & 274 & 252 \\ \hline \hline
\end{tabular}
\end{center}
\end{table}

The ratios $R^P_{\b}$ characterize the branching ratios of charmonium states
to $\gamma P$ via the virtual photon $\gamma^*$.  We summarize these ratios
and the corresponding predicted contributions to quarkonium branching
ratios in Table \ref{tab:brs}.  Here we have taken \cite{PDG} $\b(J/\psi
\to \mu^+ \mu^-) = 5.94 \times 10^{-2}$, $\b(\psi(2S) \to \mu^+ \mu^- = 7.52
\times 10^{-3}$, and $\b(\psi(3770) \to \mu^+ \mu^- = 9.71 \times 10^{-7}$.
Except for the case of $J/\psi \to \gamma \pi^0$, which we shall discuss
further in the next section, these contributions are negligible.

\begin{table}
\caption{Predicted ratios $R^P_{\b}$ and $\gamma P$ branching ratios for
quarkonium states decaying to $\gamma P$ via a virtual photon $\gamma^*$.
\label{tab:brs}}
\begin{center}
\begin{tabular}{c c c c c c c} \hline \hline
 & \multicolumn{3}{c}{$R^P_{\b}~(10^{-5})$} &
   \multicolumn{3}{c}{$\b(\gamma P)$} \\
   $P$    & $J/\psi$ & $\psi(2S)$ & $\psi(3770)$ &
 $J/\psi$ & $\psi(2S)$ & $\psi(3770)$ \\ \hline
 $\pi^0$ & 4.11 & 2.91 & 2.78 &
  $2.44 \times 10^{-6}$ & $2.19 \times 10^{-7}$ & $2.7 \times 10^{-10}$ \\
 $\eta$  & 4.45 & 3.24 & 3.10 &
  $2.64 \times 10^{-6}$ & $2.43 \times 10^{-7}$ & $3.0 \times 10^{-10}$ \\
 $\eta'$ & 4.68 & 3.62 & 3.49 &
  $2.78 \times 10^{-6}$ & $2.73 \times 10^{-7}$ & $3.4 \times 10^{-10}$ \\
\hline \hline
\end{tabular}
\end{center}
\end{table}

All the estimates presented above were for the form factor $F_{\pi^0}(Q^2)$
behaving as $2 f_\pi/Q^2$ \cite{Brodsky:1981rp}.  The calculation of Ref.\
\cite{Chernyak:1983ej} would give $R^{\pi^0}_{\b} = 10^{-4}$ at the $J/\psi$,
and hence all values approximately 2.43 times as large.
\bigskip

\centerline{\bf III.  COMPARISON WITH DATA AND VECTOR DOMINANCE}
\bigskip

The observed branching ratios \cite{PDG} for $J/\psi$ and $\psi(2S)$ decaying
to $\gamma P$ are summarized in Table \ref{tab:obs}.  The Particle Data Group
averages \cite{PDG} include
determinations by the BES Collaboration at the $J/\psi$ \cite{Ablikim:2005je}
but not the recent CLEO results \cite{CLEOGP}.  As mentioned earlier, the
single-virtual-photon process $J/\psi \to \gamma^* \to \gamma \pi^0$ is
insufficient to account for the observed branching ratio.

\begin{table}
\caption{Observed branching ratios for $[J/\psi,\psi(2S)] \to
\gamma P$ \cite{CLEOGP,PDG}, in units of $10^{-4}$.
\label{tab:obs}}
\begin{center}
\begin{tabular}{c c c c c} \hline \hline
 $P$ & \multicolumn{2}{c}{$J/\psi$} & \multicolumn{2}{c}{$\psi(2S)$} \\
 & Ref.\ \cite{CLEOGP} & Ref.\ \cite{PDG} & Ref.\ \cite{CLEOGP} 
 & Ref.\ \cite{PDG} \\ \hline
 $\pi^0$ & $0.363\pm0.036\pm0.013$ & $0.33^{+0.06}_{-0.04}$ & $<0.07$ & $<54$\\
 $\eta$  & $11.01\pm0.29\pm0.22$ & $9.8 \pm 1.0$ & $<0.02$ & $<0.9$ \\
 $\eta'$ & $52.4\pm1.2\pm1.1$ & $47.1\pm2.7$ & $1.19\pm0.08\pm0.03$
 & $1.36 \pm 0.24$ \\ \hline \hline
\end{tabular}
\end{center}
\end{table}

We now show that most of not all of the observed $\b(J/\psi \to \gamma \pi^0)$
can be accounted for by the vector dominance model (VDM).  We follow the
calculation of Ref.\ \cite{Chernyak:1983ej}, in which
\beq \label{eqn:vdm}
\frac{\Gamma(J/\psi \to \gamma \pi^0)_{\rm VDM}}{\Gamma(J/\psi \to \rho^0
\pi^0)} = \left( \frac{p^*_{\gamma \pi^0}}{p^*_{\rho \pi^0}} \right)^3
\left( \frac{e f_{\rho^0}}{m_\rho} \right)^2~.
\eeq
Here the c.m.\ 3-momenta in $J/\psi \to \gamma \pi^0$ and $J/\psi \to \rho^0
\pi^0$ are $p^*_{\gamma \pi^0} = 1546$ MeV/$c^2$ and 1448 MeV/$c^2$,
respectively.  The neutral $\rho$ meson decay constant is obtained from
that of the charged $\rho$ meson using $\tau$ decays.  Specifically
(see., e.g., Ref.\ \cite{Bhattacharya:2008ke})
\bea
f_{\rho^\pm} & = & \sqrt{2}f_\pi {\left[\frac{{\cal{B}}(\tau^- \to \nu_{\tau}\,
\rho^-)}{{\cal{B}}(\tau^- \to \nu_{\tau}\,\pi^-)}\right]}^{\frac{1}{2}}
\frac{m_{\tau}^2 - m_{\pi}^2}{m_{\tau}^2 - m_{\rho}^2}
\frac{m_{\tau}}{\sqrt{m_{\tau}^2 + 2 m_{\rho}^2}}\\
&=&(209 \pm 1.6){\rm~MeV}
\eea
where particle masses and branching fractions are taken from \cite{PDG}.  Then
\beq \label{eqn:frho}
f_{\rho^0} = f_{\rho^\pm}/\sqrt{2} = 147.8 \pm 1.1~{\rm MeV}~.
\eeq
A very similar result is obtained from the expression
\beq
\Gamma(\rho^0 \to e^+ e^-) = \frac{4 \pi \alpha^2}{3 m_\rho}f_{\rho^0}^2~.
\eeq
based on the matrix element $\langle 0 | J^{\rm em}_\mu | \rho^0(q,\epsilon)
\rangle = \epsilon_\mu m_\rho f_{\rho^0}$.  With $\b(\rho^0 \to e^+ e^-)
= (4.71 \pm 0.5) \times 10^{-5}$, $m_\rho = (775.49 \pm 0.34)$ MeV/$c^2$, and
$\Gamma_\rho=(149.4 \pm 1.0)$ MeV, one finds $f_{\rho^0}=(156 \pm 8)$ MeV.

We choose not to use the Particle Data Group average for $\b(J/\psi \to
\rho \pi)$, but to average the three most recent determinations of this
branching fraction, summarized in Table \ref{tab:rhopibr}.  Taking 1/3 of
this average for the $\rho^0 \pi^0$ final state, we have $\b(J/\psi \to
\rho^0 \pi^0) = (7.10 \pm 0.03) \times 10^{-3}$.  Eq.\ (\ref{eqn:vdm}) then
implies
\beq \label{eqn:vdmpr}
\b(J/\psi \to \gamma \pi^0)_{\rm VDM} = (28.8 \pm 1.3) \times 10^{-6}~.
\eeq
This value is consistent with the observed branching fraction quoted in
Table \ref{tab:obs}, but also allows for an additional contribution.

\begin{table}
\caption{Values of $\b(J/\psi \to \rho \pi)$ used in computing average, in
percent.
\label{tab:rhopibr}}
\begin{center}
\begin{tabular}{c c c} \hline \hline
Source & Reference & Value \\ \hline
BES direct $J/\psi$ & \cite{Bai:2004jn} & $2.184 \pm 0.005 \pm 0.201$ \\
BES $J/\psi$ from $\psi(2S)$ & \cite{Bai:2004jn} & $2.091\pm0.021\pm0.116$ \\
BaBar radiative return & \cite{Aubert:2004kj} & $2.18 \pm 0.19$ \\
Average & & $2.13 \pm 0.09$ \\ \hline \hline
\end{tabular}
\end{center}
\end{table}

In Ref.\ \cite{Chernyak:1983ej}, the relative signs of the VDM
and virtual-photon ($\gamma^*$) contributions are predicted to be positive,
so that we may place rather restricted upper bounds on the latter.  We take a
90\% confidence level (c.l.) bound of $\b(J/\psi \to \gamma \pi^0) < 41
\times 10^{-6}$.  The prediction of Ref.\ \cite{Chernyak:1983ej} was that
$\b_{\gamma^*}(J/\psi \to \gamma \pi^0)/\b(J/\psi \to e^+ e^-) = 10^{-4}$,
or (using $\b(J/\psi \to e^+ e^-) = 5.94\%$),
\beq
\b_{\gamma^*}(J/\psi \to \gamma \pi^0)_{\rm CZ} = 5.94 \times 10^{-6}~
\eeq
Let $\lambda$ be the maximum allowed fraction of the CZ branching ratio.
Then the CZ assumption of constructive interference between the VDM and
virtual-photon contributions implies
\beq
(\sqrt{28.8 \pm 1.3} + \sqrt{5.94~\lambda})^2 \le 41~,
\eeq
or, taking the lower limit of the VDM theoretical error, $\lambda < 0.23$.
The CZ continuum cross section for $e^+ e^- \to \gamma^* \to \gamma \pi^0$ at
the $J/\psi$ c.m.\ energy, corresponding to $10^{-4}$ of that for muon pair
production, is 905 fb, so the assumption of constructive interference between
VDM and virtual photon contributions in $J/\psi \to \gamma \pi^0$ implies
\beq
\sigma(e^+ e^- \to \gamma^* \to \gamma \pi^0) \le 210~{\rm fb}~,
\eeq
a bit more than half the value predicted in Table \ref{tab:K} from the form
factor in Ref.\ \cite{Brodsky:1981rp}.  The agreement between prediction and
data for $\gamma \eta^{(\prime)}$ production at 3.77 GeV suggests that not all
of the predictions of Table I are subject to the same suppression.

It is quite possible that as $|q^2| \to \infty$, the asymptotic values
of the relevant form factors are approached from below.  This is indeed
what is found for spacelike $q^2 < 0$ in singly-tagged photon-photon
collisions where one photon is highly virtual \cite{Gronberg:1997fj}.
Moreover, the corrections to the form factors in perturbative QCD are
of the form $1 - [5 \alpha_s(q^2)]/(3 \pi)$ \cite{Braaten:1982yp}, easily
entailing a suppression of the cross section by 25\% or more.
\bigskip

\centerline{\bf IV.  CONCLUSIONS}
\bigskip

We have presented continuum cross sections for $e^+ e^- \to \gamma^* \to
\gamma P$, where $P$ is a neutral pseudoscalar meson $\pi^0$, $\eta$, or
$\eta'$.  Calculations are presented at c.m.\ energies corresponding to
the masses of $J/\psi$, $\psi(2S)$, and $\psi(3770)$ for the
$\gamma^*$--$\gamma$--$\pi^0$ form factor advocated by Brodsky and Lepage
\cite{Brodsky:1981rp}, and rescaled to the $\eta$ and $\eta'$ using their
anticipated quark content and known kinematic factors.

It is shown that the contributions of the $\gamma^*$--$\gamma$--$P$ vertex
to the branching fractions for the above-mentioned charmonium states to
$\gamma P$ are negligible except in the case of $J/\psi \to \gamma \pi^0$,
where the assumption of constructive interference with the VDM contribution
permits rather stringent bounds to be placed, equivalent to continuum
cross sections for $e^+ e^- \to \gamma^* \to \gamma P$ a bit more than
half those based on the asymptotic form factor behavior predicted in Ref.\
\cite{Brodsky:1981rp}.

The cleanliness of the CLEO-c detector environment has permitted new studies
of $\gamma P$ final states at the $J/\psi$, $\psi(2S)$, and $\psi(3770)$, with
$\gamma \eta^{(\prime)}$ signals at 3.77 GeV consistent with continuum
production.
\bigskip

\centerline{\bf ACKNOWLEDGMENTS} 
\bigskip 

I thank Stan Brodsky, Brian Heltsley, Peter Kroll, Peter Lepage, and Helmut
Vogel for helpful discussions.  This work was supported in part by 
the United States Department of Energy through Grant No.\ DE-FG02-90ER-40560. 

\end{document}